\documentclass[conference,letterpaper]{IEEEtran}
\usepackage{cite}      % Written by Donald Arseneau

\usepackage{stfloats}  % Written by Sigitas Tolusis
\usepackage[latin1]{inputenc}
\usepackage[T1]{fontenc}
\usepackage{textcomp}
\usepackage{subfigure}
\usepackage{amssymb,amsmath,xspace,tabularx}
\providecommand{\href}[2]{#2} 
\providecommand{\hypersetup}[1]{}\providecommand{\url}[1]{#1}
  \usepackage[dvips]{graphicx} % um .eps Dateien einzubinden 
  \usepackage[hypertex]{hyperref} 
\begin{document}

\title{On the deployment of Mobile Trusted Modules}

\author{\authorblockN{Andreas U.\ Schmidt,}
\authorblockN{Nicolai Kuntze, Michael Kasper}
\authorblockA{Fraunhofer Institute for Secure Information Technology SIT\\
Rheinstra\ss e 75, 64295 Darmstadt, Germany\\
Email: \{andreas.schmidt,nicolai.kuntze,michael.kasper\}@sit.fraunhofer.de}}

% avoiding spaces at the end of the author lines is not a problem with
% conference papers because we don't use \thanks or \IEEEmembership

%Homer Simpson\authorrefmark{2},
%Montgomery Scott\authorrefmark{3} and
%Eldon Tyrell\authorrefmark{4}}
%Telephone: (800) 555--1212, Fax: (888) 555--1212}
%\authorblockA{\authorrefmark{4}Tyrell Inc., 123 Replicant Street, Los Angeles, California 90210--4321}}

%\specialpapernotice{(Invited Paper)}

\maketitle

\begin{abstract}
In its recently published \textit{TCG Mobile Reference Architecture}, 
the TCG Mobile Phone Work Group specifies a new concept to enable 
trust into future mobile devices. 
For this purpose, the TCG devises a trusted mobile platform as a 
set of trusted engines on behalf of different stakeholders 
supported by a physical trust-anchor. 
In this paper, we present our perception on this emerging specification. 
We propose an approach for the practical design and implementation of this concept 
and how to deploy it to a trustworthy operating platform.
In particular we propose a method for the take-ownership of a device by the
user and  the migration (i.e., portability) of user credentials between devices.
\end{abstract}


% for peerreview papers, inserts a page break and creates the second title.
% Will be ignored for other modes.
\IEEEpeerreviewmaketitle

\section{Introduction}
As a first deliverable within their scope and work programme, 
the Mobile Phone Work Group of the Trusted Computing Group (TCG MPWG) 
has published a specification~\cite{TCG_MPWG_Specification}, 
which offers new potentials for implementing trust in mobile computing platforms 
by introducing a new, hardware-based trust anchor for mobile phones and devices. 
This trust anchor, called a Mobile Trusted Module (MTM), 
has properties and features comparable to a Trusted Platform Module 
(TPM, see \cite{TCG_Architecture,TCG_TPM_Specification}). 
Concurrently the MPWG issued a much more universal security architecture 
for mobile phones and devices on a higher abstraction level.  
The pertinent specification is called 
\textit{TCG Mobile Reference Architecture (RA)}~\cite{TCG_MPWG_Architecture} 
and abstracts a trusted mobile platform as a set of tamper resistant trusted 
engines operating on behalf of different stakeholders. 
This architecture offers a high degree on flexibility and modularity 
in design and implementation of the trusted components to all 
participants in hard- and software development. 

An important aspect of the \textit{TCG Mobile Reference Architecture} is the potential to virtualise 
significant parts of a trusted mobile platform as trusted software applications and services. 
The trusted execution chain for this rests on the MTM. 
The implementation of this chip depends on the security requirements of 
its specific use-case. 
For high levels of protection and isolation, an MTM 
could be implemented as a slightly modified Trusted Platform Module (TPM). 
This enables cost-effective implementation of new security-critical 
applications and various innovative business models, in both the mobile and 
generic computing domain~\cite{KuntzeSchmidt2006A,KuntzeSchmidt2006C,KuntzeSchmidt2007B}.

The present paper discusses the main structural features of the RA, highlighting
the  capabilities of the MTM as the main functional building block.
After this technology review, we propose two basic methods for usage of the RA,
namely the set-up of a trusted subsystem on a device by a remote owner, and
its migration from one device to another.

This paper is organised as follows.
 In Section \ref{section:tcg_mpwg_architecture}, 
we explore the significant parts of the 
\textit{MPWG Reference Architecture}. 
It is divided into four parts. 
Subsection \ref{section:tcg_mpwg_architectural_overview} 
gives an overview of the security architecture, and 
subsection \ref{section:mobile_trusted_module} details the concepts of the proposed
architectural approach for an MTM and the requirements to virtualise its functionality, 
whereby a high security and isolation level is maintained. 
Furthermore, we propose a model for remote stakeholder take-ownership 
in \ref{section:ro_takeownership} and 
migration of trusted subsystems in 
\ref{section:migration_remote_stakeholder_subsystems}. 
In Section \ref{section:design_trusted_engines}, 
we show how such an architecture can be implemented on trustworthy operating platforms. 
\section{TCG MPWG Reference Architecture}\label{section:tcg_mpwg_architecture}
The TCG MPWG has developed an architecture on a high 
level of abstraction for a trusted mobile platform, 
which offers numerous variations for design and implementation. 
In this section, we reflect  essential parts of this 
architecture and an overview of significant platform 
components in terms of our objective.
\subsection{Architectural Overview}\label{section:tcg_mpwg_architectural_overview}
\begin{figure}[ht]
 	\centering
 	\includegraphics[width=1\columnwidth]{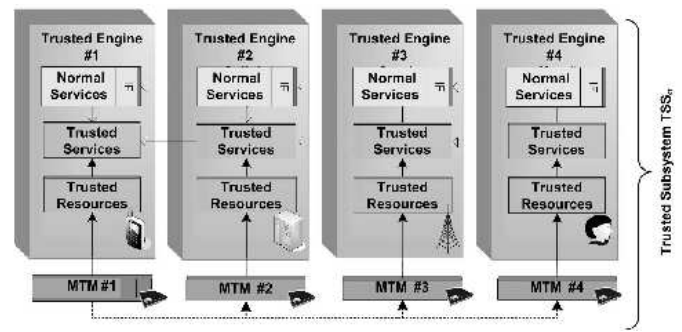}
 	\caption{Trusted Mobile Platform Architecture}
 	\label{fig:tcg_mpwg_trusted_mopile_platform_architecture}
 \end{figure}

A trusted mobile platform is characterised as a set of multiple tamper-resistant engines, 
each acting on behalf of a different stakeholder. 
Broadly, such an platform has several major components: 
trusted engines $\mathcal{TE}$, trusted services $\mathcal{TS}$ customised by trusted 
resources $\mathcal{TR}$. 
A general trusted mobile platform is 
illustrated in Figure \ref{fig:tcg_mpwg_trusted_mopile_platform_architecture}. 

We define a trusted subsystem $\mathcal {TSS}_{\sigma}$ as a logical unit 
of a trusted engine together with its interrelated hardware compartment. 
A $\mathcal {TSS}$ of a stakeholder $\sigma$ can formally described by a tuple 
\begin{eqnarray} \nonumber
 \mathcal {TSS_\sigma} & = & \left\lbrace 
				\mathcal{TE_\sigma}, \mathcal{TS_\sigma}, \mathcal{TS_\epsilon},
				\mathcal{TR_\sigma}, \mathcal{SP_\sigma},
				\mathcal{SC_\sigma}
				\right\rbrace
\end{eqnarray}
In each trusted subsystem $\mathcal {TSS}$ 
either a remote or local entity acts as a stakeholder, 
who is able to configure its own subsystem and define 
his security policy $\mathcal{SP_\sigma}$ and system configuration $\mathcal{SC_\sigma}$ 
within an isolated and protected environment. 
The \textit{MPWG Reference Architecture} 
specifies the following principal entities: 
the local stakeholders \textit{Device Owner $\mathcal{DO}$} 
and \textit{User $\mathcal{U}$}; and the remote stakeholders \textit{Device Manufacturer $\mathcal{DM}$}, 
and more general \textit{Remote Owners $\mathcal{RO}$} (e.g.\ a communication carrier, or service provider).
The functionality of a $\mathcal {TSS}$ is either based on dedicated resources 
of an embedded engine $\mathcal{TE_\sigma}$, or may be provided by 
trusted services $\mathcal{TS_\epsilon}$ of external engines. 

Each subsystem is able to enforce its security policy $\mathcal{SP_\sigma}$ and 
subsystem configuration $\mathcal{SC_\sigma}$. As a consequence, the functionality of a
trusted subsystem $\mathcal {TSS_\sigma}$ is constrained by the available 
resources $\mathcal{TR_\sigma}$ with their derived trusted services 
$\mathcal{TS_\sigma}$, by the offered functionality of external 
trusted services $\mathcal{TS_\epsilon}$, by the security policy 
$\mathcal{SP_\sigma}$,  and finally the system configuration $\mathcal{SC_\sigma}$ 
of an engine's stakeholder. 

All internal functions executed inside $\mathcal {TSS}_\sigma$ 
are isolated from other subsystems by the underlying security 
layer and is only accessible if a proper service interface 
is defined and exported. 
A $\mathcal {TSS}_\sigma$ relies on the reputation of the 
stakeholder $\sigma$ as basis for that trust. Therefore, 
each stakeholder issues a security policy $\mathcal{SP_\sigma}$ 
and a set of credentials belonging to embedded trusted 
components of its subsystem $\mathcal {TSS}_\sigma$. 
This policy contains reference measurements (RIM), 
quality assertions and security-critical requirements. 

\subsubsection{Trusted Engines}\label{section:trusted_engines}
The most important concept within the \textit{MPWG Reference Architecture} is that of trusted engines. 
\begin{figure}[ht]
	\centering
	\includegraphics[width=0.90\columnwidth]{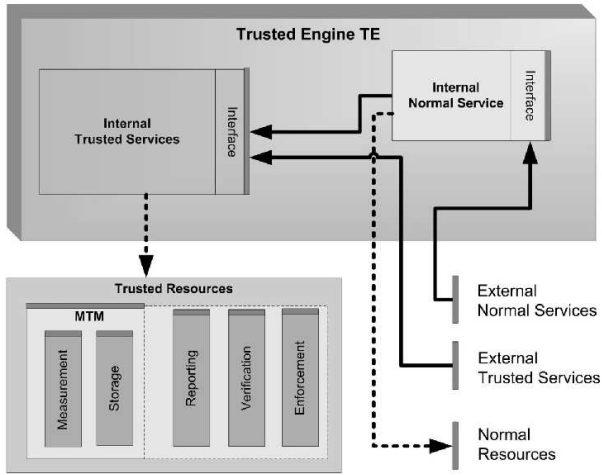}
	\caption{Generic Trusted Engine}
	\label{fig:tcg_mpwg_generic_engine}
\end{figure}
The purpose of a trusted engine is to provide confidence in all its embedded services, 
which are internally or externally provided by the engine. 
It is a protected entity on behalf of a specific stakeholder 
that has special abilities to manipulate and store data, 
and to provide evidence of its trustworthiness and the current state of the engine. 
Figure \ref{fig:tcg_mpwg_generic_engine} shows a generic trusted engine. 
In general, each engine has at least following abilities:
\begin{itemize}
 \item implement arbitrary software functionalities as trusted and/or normal services,
 \item provide the evidence for its trustworthiness,
 \item report the evidence of its current state,
 \item obtain and use Endorsement Keys (EK) and/or Attestation Identity Keys (AIK),
 \item access a set of trusted resources, and
 \item import and/or export services, shielded capabilities and protected functionality.
\end{itemize}

In order to establish a definite categorisation, the MPWG differentiates engines 
according to their functional dispensability. Therefore, an engine is either 
dedicated to a mandatory (of $\mathcal{DO}$ or $\mathcal{DM}$) or a 
discretionary domain (of $\mathcal{DO}$). 
Engines inside a mandatory domain are permanently located on a trusted platform 
and hold security-critical and essential functionality. 
All essential services of a trusted mobile platform should be located inside the 
mandatory domain, which does \emph{not permit a local stakeholder to remove a 
remote owner from the engine}. 
Mandatory engines have access to a \textit{Mobile Remote owner Trusted Module (MRTM)} 
to guarantee that a valid and trustworthy engine state is always present.

Non-essential engines and services are replaceable by the device owner 
$\mathcal{DO}$ and should be located inside the discretionary domain. 
Engines inside the discretionary domain are controlled by the device owner 
$\mathcal{DO}$. 
Discretionary engines are required to be 
supported by a \textit{Mobile Local owner Trusted Module (MLTM)}.

\subsubsection{Trusted Resources} \label{section:trusted_resources}
As illustrated in Figure \ref{fig:tcg_mpwg_generic_engine}, an
internal trusted service has access to several trusted resources. 
The TCG calls these resources \textit{Root-Of-Trusts (RoT)} representing 
the trusted components acting on base of the trusted execution chain 
and providing functionality for measurement, storing, reporting, 
verification and enforcement that affect the trustworthiness of the platform. 
The following RoTs are defined for the mobile domain:
 \begin{itemize}
 \item Root of Trust for Storage (RTS),
 \item Root of Trust for Reporting (RTR),
 \item Root of Trust for Measurement (RTM),
 \item Root of Trust for Verification (RTV), and
 \item Root of Trust for Enforcement (RTE) 
\end{itemize}
Each RoT vouches its trustworthiness either directly by 
supplied secrets (EK, AIK) and associated credentials, 
which are only accessible by authenticated 
subjects of the stakeholder, or indirectly by measurements 
of other trusted resources. 
These resources are only mutable by authorised entities of a stakeholder. 

In this paper, we group several logically self-contained RoTs 
to simplify the presentation of interfaces and the communication layer. 
In a typical arrangement, the RTS and RTR represent one unit, while the 
RTM and RTV build another unit within an $\mathcal{TSS}_\sigma$. 
However, note that the RTV and the RTM depend on 
protected storage mechanisms, which are provided by the RTS. 
Thus, it is also plausible to implement all RoTs together as a common unit within an engine.

\textbf{RTS/RTR}
are the trusted resources that are responsible for secure storage 
and reliable reporting of information about the state of trusted mobile platform.  
An RTS provide PCRs and protected storage for an engine and stores the measurements made by the RTM, 
cryptographic keys, and security sensitive data. An RTR signs the measurements with cryptographic 
signature keys of $TSS_\sigma$.

\begin{figure}[t]
	\centering
	\includegraphics[width=0.90\columnwidth]{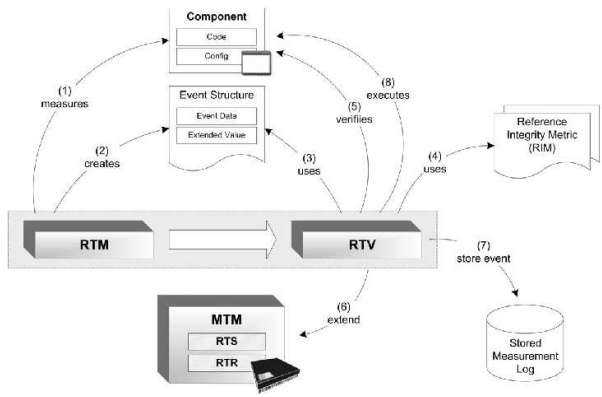}
	\caption{Measurement and Verification Process}
	\label{fig:tcg_mpwg_measurement_and_verification}
\end{figure}

\textbf{RTM/RTV} In general, an RTM is a reliable instance to measure software components and 
provide evidence of the current state of a trusted engine and its embedded services. 
In the mobile domain, to avoid communication costs, this functionality is extended by a 
local verifier, which checks the measurements 
against a given \textit{Reference Integrity Metrics} (RIM). 
This process can be done instantly as the measurements are performed employing 
a combination of RTM and RTV. 
Figure  \ref{fig:tcg_mpwg_measurement_and_verification} depicts such a \textit{Measure$\to$Verify$\to$Extend} process.

An \textbf{RTE} is required if an engine uses allocated resources and services. 
In this case, such RoT acts as a trusted boot loader and ensures the availability 
of all allocated trusted resources and services within that trusted subsystem. 

\subsubsection{Services of a Trusted Engine} A trusted engine integrates all functionality by customising 
available platform resources as software services. Such a service offers computation, storage, or communication 
channels to other internal or external services and applications based on dedicated or allocated resources. 
The MPWG categorises them into: trusted, normal, and measured services. 

A trusted service customises trusted resources. 
Thus, a trusted service is implicitly supplied with an $EK$ or $AIK$ in order to attest its trustworthiness. 
Trusted services are intended to provide reliable measurements of their current state and to provide evidence 
of the state of other normal services or resources.

Normal services are customising normal resources and implement functionality, 
but are not able to provide evidence of their trustworthiness by own capabilities. 
However, normal services can access internal trusted services to use their provided 
functionality. Therefore, an internal normal services is able to vouch its 
trustworthiness by associated integrity metrics that have been measured by a trusted service.

\subsection{Mobile Trusted Module} \label{section:mobile_trusted_module}
The generic term \textit{Mobile Trusted Module (MTM)} refers to a dedicated hardware-based trust-anchor. 
It is typically composed of an RTS and RTR and has characteristics comparable to a TPM. 
According to their design objective the MPWG distinguishes between MRTM and MLTM. 
Both must support a subset of TPM commands as specified in~\cite{TCG_MPWG_Architecture}. 
Additionally, an MRTM has to support a set of commands to enable local verification and specific mobile device functionality. 

The \textit{TCG MPWG Reference Architecture} does not exclude to utilise a 
TPM v1.2 (or even a TPM v1.1) as an MTM, if an appropriate interface consisting of a set of 
commands conforming to the MPWG specification and associated data structures are provided. 
Although it is possible to implement this architecture upon a standard TPM, 
we here focus on a more sophisticated solution based on a Trustworthy Computing Platform such 
as EMSCB/Turaya~\cite{EMSCB}. 
In this context, we expect three different solutions for isolation,
key management and protection of $\mathcal{TSS}_\sigma$ . 

A \textbf{Standard TPM-based Model} uses a non-modified standard TPM to build the 
trusted computing base. The secret keys are stored into a single key-hierarchy on 
behalf of $\mathcal{DO}$ as specified in \cite{TCG_MPWG_Specification}. 
In this case, an adversary or malicious local owner may be able to access the secret 
keys of a remote stakeholder and take control of a remote owner compartment. 
A $\mathcal{DO}$ can also disable the whole MTM or corrupt mandatory engines of remote stakeholders.

A \textbf{Software-based MTM-Emulation Model} uses a software-based allocated $MTM$-emulation with 
an isolated key-hierarchy. All sensitive and security-critical, such as $EK$ or $SRK$, are only 
protected by software mechanisms outside of the tamper-resistant environment of a dedicated 
MTM \cite{vtpm_ibm_xen,TPM_Emulator}.

\textbf{Generic MTM-based Model supporting multiple stakeholders and virtual MTMs.} 
In order to circumvent resulting drawbacks and mitigate attacks, we favour a solution with 
a higher level of security.  For this reason, we adopt the proposed secure co-processor 
variant of~\cite{vtpm_ibm_xen} and describe a generic MTM with support for multiple 
stakeholder environments.
\begin{figure}[ht]
	\centering
	\includegraphics[width=\columnwidth]{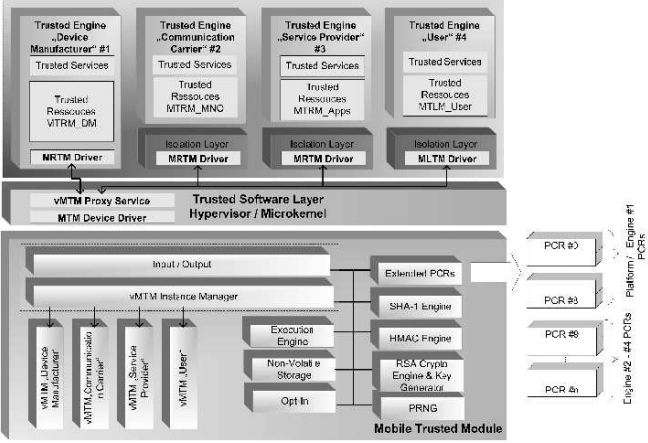}
	\caption{MTM Architecture supporting multiple Stakeholders}
	\label{fig:tcg_mpwg_vmtm_virtualization}
\end{figure}
In a cost-efficient scenario, the trusted mobile platform is implementable based on a 
single generic MTM and several virtualised MTMs for each trusted engine. Hence, at least 
one dedicated MTM has to be available and additionally a unique vMTM has to be instantiated 
in each trusted subsystem $\mathcal {TS_\sigma}$. In such case, a physically bounded MTM in the platform 
acts as a master trust anchor and offers MRTM and MLTM functionality with respect to its specific use case. 

A Trusted Software Layer offers a \textit{vMTM Proxy Service} to all embedded trusted engines 
$\mathcal{TE}_\sigma$. The main task of this service is to route MTM commands from a $\mathcal{TE}_\sigma$ to 
its dedicated instance $vMTM_\sigma$. The advantage is that all security-critical MTM commands are tunnelled to 
$vMTM_\sigma$ and are executed within the protected environment of the dedicated MTM.

Figure \ref{fig:tcg_mpwg_vmtm_virtualization} illustrates the architecture of a generic MTM with isolated 
vMTM compartments. This architecture requires a slightly modified TPM. Mainly, we add a trusted component, 
the \textit{vMTM Instance Manager}, which is responsible to separate vMTM instances from each other. 
This includes administration, isolated execution, memory management and access control for each stakeholder 
compartment. Thus, a vMTM instance is able to hold an autonomous and hardware-protected key hierarchy to store 
its secrets and protect the execution of security-critical data (e.g.\ signature and encryption algorithms).
\subsection{Setup and Take-Ownership of a Trusted Subsystem} \label{section:ro_takeownership}
\begin{figure}[ht]
	\centering
	\includegraphics[width=\columnwidth]{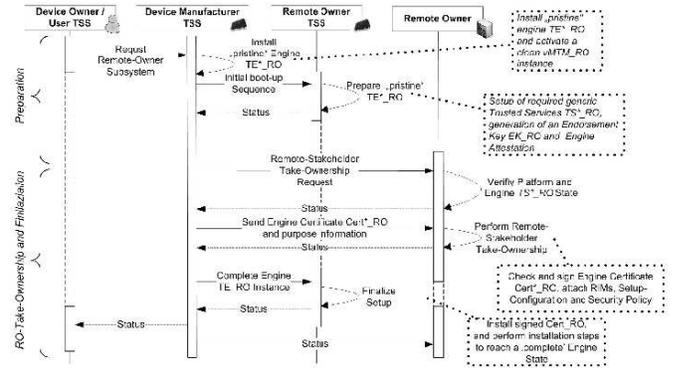}
	\caption{Remote Stakeholder Take-Ownership Protocol}
	\label{fig:tcg_mpwg_vmtm_takeownership_sequence}
\end{figure}

The take-ownership operation establishes  the trust relationship between a stakeholder and trusted mobile platform. 
Currently, the \textit{MPWG Reference Architecture} does not define how a remote owner is to perform this initial 
setup and take-ownership of its $\mathcal{TSS_\sigma}$. Hence we propose a method in this section.
The main idea behind our procedure is to install and instantiate a 'blank' trusted subsystem 
$\mathcal {TSS^*_\sigma}$ containing a pristine engine $\mathcal{TE}^*_\mathcal{RO}$ with a set of 
generic trusted services $\mathcal {TS^*_\sigma}$. This subsystem is then certified by a remote owner, 
if the platform is able to provide evidence of its pristine configuration and policy conformance 
with respect to $\mathcal{RO}$. Figure~\ref{fig:tcg_mpwg_vmtm_takeownership_sequence} illustrates this process,
which we now descirbe. 

\textbf{Platform and Protocol Precondition.} In a preliminary stage, 
the trusted mobile platform has carried out the boot process and has loaded the trusted computing 
base and the engine $\mathcal{TE}_\mathcal{DM}$ with its trusted services. 
The trusted platform has checked that  the installed  hardware and running software are in a 
trustworthy state and configuration. It is able to report and attest this state, if 
challenged by an authorised entity.

\textbf{Remote Stakeholder Take-Ownership Protocol.} In the first phase, the trusted engine 
$\mathcal{TE}_\mathcal{DM}$ carries out a take-ownership preparation for the remote stakeholder. 
A 'blank' engine $\mathcal{TE}^*_\mathcal{RO}$ is installed and booted by the $RTE_\mathcal{DM}$, 
and a clean $vMTM_\mathcal{RO}$ instance is activated inside the dedicated $MTM$. 
An initial setup prepares the pristine engine $\mathcal{TE}^*_\mathcal{RO}$. 
A endorsement key-pair $EK^*_\mathcal{RO}$ is generated within $vMTM_\mathcal{RO}$ with a 
corresponding certificate $Cert_\mathcal{TSS_\mathcal{RO}}$\footnote{Typically, the key generation 
needs a so-called \textit{Owner Authentication}. Because this is problematic in a remote-owner scenario, 
authentication of command execution may be enforced by challenge-response mechanisms 
between $\mathcal{RO}$ and $\mathcal{TSS}_\mathcal{RO}$.}.

Next, $\mathcal{TE}^*_\mathcal{RO}$ performs an attestation of its current state. 
The attestation can be done by the local verifier $RTV_\mathcal{DM}$ inside the $\mathcal{TSS}_{DM}$ 
using $RIM$ certificates of the remote stakeholder $\mathcal{RO}$. 
If no suitable $RIM$ and corresponding $RIM$-certificate are available for an pristine engine, 
alternatively a remote attestation with an associated Privacy CA is also possible.

$\mathcal{TE}^*_\mathcal{RO}$ creates a symmetric key $K_{\mathcal{RO},temp}$ and encrypts the 
public portion of the endorsement key $EK^*_\mathcal{RO}$, the corresponding certificate $Cert_\mathcal{TSS_\mathcal{RO}}$, 
attestation and purpose information. Next, $\mathcal{TE}^*_\mathcal{RO}$ encrypts 
$K_{\mathcal{RO},temp}$ with a public key $K_{\mathcal{RO},PK}$ and sends both messages to the remote owner.
After reception by the remote stakeholder, the messages are decrypted using the private portion of 
key $K_{\mathcal{RO},PK}$. We assume that this key is either available through a protected communication channel 
or pre-installed by the device manufacturer.

In a next step, $\mathcal{RO}$ verifies the attestation data and checks the intended 
purpose of $\mathcal{TE}^*_\mathcal{RO}$. If the engine and device attestation data is 
valid and the intended purpose is acceptable, the $\mathcal{RO}$ generates an individual 
security policy $\mathcal{SP}_{RO}$. The $\mathcal{RO}$ signs the $Cert_\mathcal{TSS_\mathcal{RO}}$ 
and creates $RIM$ certificates for local verification of a 'complete' $\mathcal{TSS}_\sigma$. 
Furthermore, $\mathcal{RO}$ creates a Setup Configuration $SC_\mathcal{TSS_\mathcal{RO}}$, 
which enforces the engine to individualise its services and complete its configuration with 
respect to the intended purpose and given security policy. 
In this step, $\mathcal{RO}$ encrypts the messages with the public portion of the 
$K_{\mathcal{RO},EK}$ and transfers this package to the engine $\mathcal{TE}_{RO}$. 

Finally, the trusted engine 
$\mathcal{TE}^*_\mathcal{RO}$ decrypts the received package and installs it inside the 
$TSS_\mathcal{RO}$ and thus completes its instantiation.
\subsection{Migration of a Trusted-Subsystem} \label{section:migration_remote_stakeholder_subsystems}
If a stakeholder wants to move a source $\mathcal{TSS}_{\sigma,S}$ to another MTM-enabled platform, for instance to
port user credentials from device to device, all security-critical information including the Storage Root Key 
(SRK) has to be migrated to the target $\mathcal{TSS}_{\sigma,D}$. 
In our scenario, we assume the same remote owner (e.g.\ mobile network operator)
on both subsystems $\mathcal{TSS}_{RO,S}$ and $\mathcal{TSS}_{RO,D}$. 
\begin{figure}[ht]
	\centering
	\includegraphics[width=\columnwidth]{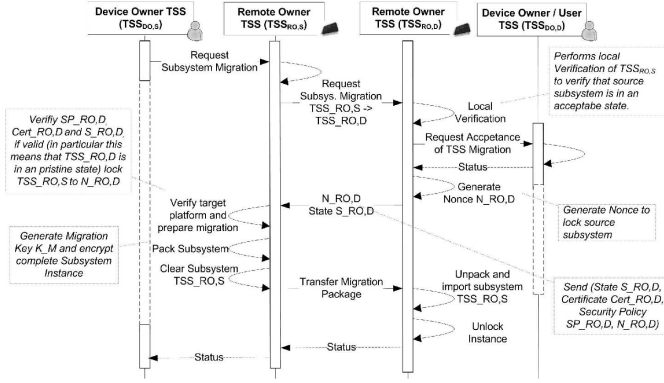}
	\caption{Trusted Subsystem Migration Protocol}
	\label{fig:tcg_mpwg_vmtm_migration_protocol}
\end{figure}

To be able to securely migrate the SRK, we suggest a modification of the current MPWG specification 
to allow \textit{inter-stakeholder-migration} of a complete isolated key hierarchy. 
Thus, an isolated key hierarchy is (1) migratable between environments of identical stakeholders, 
(2) if and only if an entitling security policy on both platforms exists. 
The advantage of migration between identical stakeholder subsystems is that the migration process 
doesn't require a trusted third party. We only involve the owner in combination with local verification 
mechanisms of the $\mathcal{TSS}_{RO}$ to migrate the trusted subsystem (including the SRK) to another platform. 
This enables for instance direct, device-to-device porting of credentials, e.g.\ using short-range communication.
We here propose a complete, multilateral and secure migration protocol, which is 
illustrated in Figure~\ref{fig:tcg_mpwg_vmtm_migration_protocol}.

\textbf{Platform and Protocol Precondition.} 
Similar to section \ref{section:ro_takeownership}, the trusted mobile platform has carried out 
the same initial steps as mentioned above. Furthermore, the remote owner has performed an remote 
take-ownership procedure as described in \ref{section:ro_takeownership}.

\textbf{Trusted Subsystem Migration Protocol.} 
At the beginning of the migration protocol, the device owner $\mathcal{DO}_S$ of the source 
platform $\mathcal{TP}_S$ initialises the migration procedure and requests an appropriate migration 
service of $\mathcal{TSS}_{RO,S}$. Next, the trusted platform $\mathcal{TP}_S$ is instructed by 
$\mathcal{TSS}_{RO}$ to establish a secure channel to the target platform $\mathcal{TP}_D$.
After the connection is available, $\mathcal{TSS}_{RO,S}$ activates the corresponding migration 
service of $\mathcal{TSS}_{RO,D}$ to perform the import procedure. 
Thereupon, the target subsystem $\mathcal{TSS}_{\sigma,D}$ performs a local verification of 
$\mathcal{TSS}_{RO,S}$. If revoked, it replies with an error message and halts the protocol. 
Otherwise $\mathcal{TE}_{RO,D}$ requests an confirmation from the local owner $\mathcal{DO}_D$. 

Next, the target subsystem $\mathcal{TSS}_{RO,D}$ generates a nonce $N_{RO,D}$. In order to provide 
evidence of its trustworthiness, $\mathcal{TSS}_{RO,D}$ sends all necessary information to the source 
subsystem $\mathcal{TSS}_{RO,S}$. This includes the current state $S_{RO,D}$, a certificate of 
$\mathcal{TSS}_{RO,D}$, security policy $\mathcal{SP}_{RO,D}$ and the nonce $N_{RO,D}$.
Having received the target subsystem's message, $\mathcal{TSS}_{RO,S}$ verifies the state of 
$\mathcal{TSS}_{RO,D}$. If the target system is in a trustworthy state and holds an acceptable security 
policy and system configuration, the current state of $\mathcal{TSS}_{RO,S}$ is locked to nonce $N_{RO,D}$. 

The $\mathcal{TSS}_{RO,S}$ generates a symmetric migration key $K_M$, serialises its instance and encrypts 
it with the migration key, which is  bound to an acceptable configuration of $\mathcal{TSS}_{RO,D}$. Next, 
the key-blob and the encrypted instance are sent to the destination $\mathcal{TSS}_{RO,D}$. In particular, 
this includes the whole isolated key-hierarchy $\mathcal{K}_{\mathcal{RO},S}$ with $SRK_{\mathcal{RO},S}$, 
the security policy $\mathcal{SP}_{RO,S}$, and the required subsystem configuration $\mathcal{SC}_{RO,S}$.

Finally, the target subsystem $\mathcal{TSS}_{RO,D}$ decrypts the received blob and uses $SRK_{RO,S}$ as 
its own $SRK$. The subsystem verifies the obtained security policy $\mathcal{SP}_{RO,S}$ and the subsystem 
configuration $\mathcal{SC}_{RO,S}$.  With this information, $\mathcal{TSS}_{RO,D}$ rebuilds the 
internal structure of the source.

The source system should then be notified of the success of migration and 
ultimately delete the migrated key hierarchy (or even do it before sending the migration package as
indicated for simplicity in Figure~\ref{fig:tcg_mpwg_vmtm_migration_protocol}). 
Otherwise one obtains replicated trusted subsystems, by themselves indistinguishable to the remote owner.
But this may depend on the policies to be enforced in the particular use case.
\begin{figure}[ht]
	\centering
	\includegraphics[width=1\columnwidth]{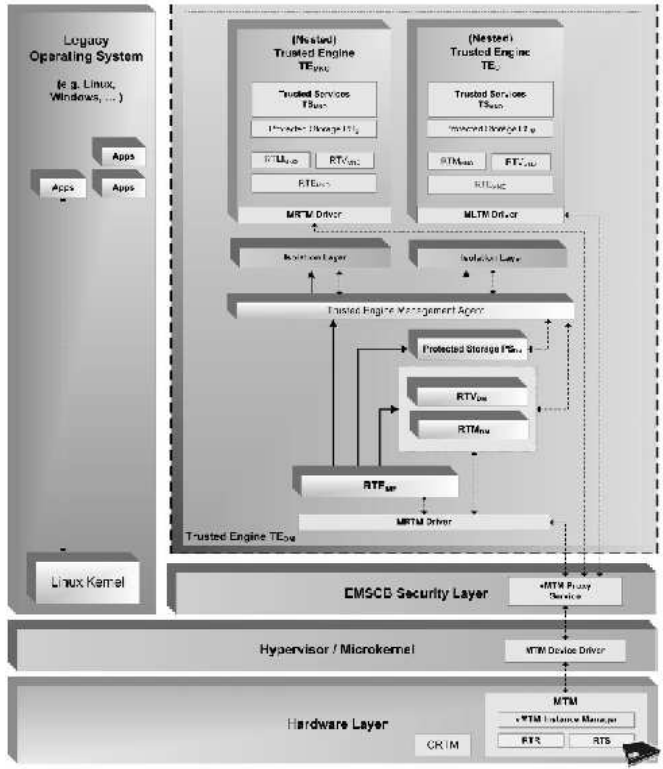}
	\caption{Trustworthy Operating Platform with multiple Trusted Engines}
	\label{fig:mtm_vsim_emscb_platform}
\end{figure}

\section{Design of Mobile Trusted Modules on Trustworthy 
Operating Platforms}
\label{section:design_trusted_engines}

A prototypical implementation of the trusted engines and the specified 
trusted services was realised as an extension to the existing EMSCB / Turaya 
Computing Platform. 
Turaya is an implementation of the EMSCB security architecture. It provides 
fundamental security mechanisms and a protected and isolated execution 
environment, which meet the requirements of the \textit{MPWG Reference 
Architecture}~\cite{EMSCB,EMSCB_01,PerseusOS}. 

Figure~\ref{fig:mtm_vsim_emscb_platform} illustrates our model, in which a 
hypervisor/microkernel executes a legacy operating system in coexistence with 
a running instance of the EMSCB-based security architecture. The latter 
controls a virtual machine with several trusted engines and services 
compliant to the MPWG requirements~\cite{TCG_MPWG_Architecture,TCG_MPWG_Specification}. 
In the following paragraphs, we outline  the significant platform layers concerning our 
approach.

The \textbf{Hardware Layer} of our model
includes a generic MTM as described in Section~\ref{section:mobile_trusted_module}, 
in addition to  conventional hardware components. This MTM acts as a 
dedicated master trust anchor for the complete trusted mobile platform. 

The \textbf{Virtualisation Layer} provides generic hardware abstraction, 
between the physical hardware of a  trusted mobile platform and the 
\textit{Trusted Software Layer} below. The EMSCB 
project supports microkernels of the L4-family~\cite{Drops} such 
as hypervisors~\cite{barham03xen}. In general, all solutions provide mechanisms for 
resource management, inter-process-communication (IPC), virtual machines, 
memory management and scheduling. In our specific case, the virtualisation 
layer includes also a fully functional device driver for a dedicated generic 
MTM. Furthermore, it is responsible for instantiation of both the trusted 
software layer and the legacy operating system.

The \textbf{Trusted Software Layer} 
provides security functionality and is responsible for isolation of embedded 
applications and software compartments. It also implements the \textit{vMTM 
Proxy Service} as described in Section~\ref{section:mobile_trusted_module}. 
Currently, EMSCB/Turaya provide an excellent foundation by its security 
services (trust manager, compartment manager, storage manager), which are 
required by the RTR and RTV, Protected Storage and Trusted Engines Management 
Agent of $\mathcal{TE}_{DM}$. Therefore, it is reasonable to build the 
significant parts of the device manufacturer engine $\mathcal{TE}_{DM}$ 
within this layer.

 Trusted engines $\mathcal{TE}_\sigma$ within 
the \textbf{Application Layer} are implemented as parallel and isolated 
L4Linux compartments~\cite{L4Linux} on behalf of different stakeholders. Each 
compartment has access to its vMTM instance through an embedded client-side 
device driver. This driver constrains the functionality with respect to its specific 
use case (MRTLM or MLTM). Furthermore, $\mathcal{TE}_\sigma$ has an 
$RTE_\sigma$, which is responsible for building all required allocated 
trusted resources and services depending of its specific system configuration 
$\mathcal{SC}_\sigma$ and security policy $\mathcal{SP}_\sigma$. 
\section{Conclusion and Further Work}
We have introduced the Trusted Engines and MTMs in terms of our objective. 
In this context, we have exposed significant parts of the MPWG Reference Architecture and how it can be
implemented on a (very slightly modified) TPM trust-anchor. 
We have shown how to deploy trusted virtualised compartments on devices and
exhibited basic operations required in the mobile domain, such as migration.

Using a vMTM in lieu of a Subscriber Identity Module (SIM) 
as a trusted and protected software allows expansion to a much wider field of 
authentication and identification management systems even 
on standard PC platforms~\cite{SIMTrustParameters}. 
Supporting online transactions by  authentication via credentials held in a vMTM 
may be one attractive use case. 
However, there are some privacy and security challenges associated with 
this implementation on a desktop computer, which require further research.
Finally, replacing SIMs/USIMs by multi-purpose vSIMs may be attractive even
for genuine mobile devices.

%\bibliographystyle{unsrt}
%\bibliography{bibliography}

 \providecommand{\noopsort}[1]{} \providecommand{\singleletter}[1]{#1}

\end{document}